\documentclass[12pt]{amsart}
\usepackage{amsaddr}
\usepackage[noshift]{marktext}
\usepackage[]{svmacro}
\newcommand{\AB}[2][]{\sidenote[colback=teal!10]{\textbf{AB\xspace #1:} #2}}

\usepackage{algorithm}
\usepackage{algpseudocode}
\usepackage{bm} 
\numberwithin{equation}{section} 

\hypersetup{
	colorlinks,
	linkcolor={red!50!black},
	citecolor={green!50!black},
	urlcolor={blue!80!black}
}

\usepackage{enumitem} 
\usepackage{mhchem} 
\usepackage{lipsum}
\usepackage{metalogo} 
\usepackage{subcaption} 
\usepackage{graphicx} 
\usepackage{cancel}

\usepackage[style=alphabetic, backend=biber]{biblatex}
\title{Self-construction and destruction of living transport networks}
\author{Chanoknun Sintavanuruk and Yoichiro Mori}
\address{Department of Mathematics, University of Pennsylvania, Philadelphia PA, 19104, USA}
\author[Farhadifar]{Reza Farhadifar}
\address{Center for Computational Biology, Flatiron Institute, New York, NY 10010, USA}
\author[Katifori]{Eleni Katifori}
\address{Department of Physics and Astronomy, University of Pennsylvania, Philadelphia PA, 19104, USA}
\address{Center for Computational Biology, Flatiron Institute, New York, NY 10010, USA}

\begin{document}
\sloppy 
\maketitle

\begin{abstract}
Biological transport networks
adapt through dynamic interactions between material transport and structural modification during growth and development.
In this work, we present a model of transport network growth driven by local material concentration.
Using an advection-diffusion framework on a metric graph with a tip growth rule, we investigate how successive construction and destruction influence network development.
Our results reveal that while repeated cycles of elongation and retraction can facilitate growth enhancement, a network need to have a structure that mitigates material dissipation.
This finding suggests that additional regulatory mechanisms are necessary for networks to efficiently redistribute resources following structural retraction.
\end{abstract}

\tableofcontents

\section{Introduction}

Transport networks are ubiquitous in biological systems, manifesting in structures such as the tree-like morphology of neurons, leaf venation patterns \cite{katifori_damage_2010}, the cardiovascular system, and the intracellular networks of slime molds.
These networks have been optimized over millions of years of evolution to perform essential transport functions that sustain life.
A shared characteristic of such networks is their ability to self-construct, a process that involves the relocation and consumption of mass.
Remarkably, this self-construction often depends on the very transport network being formed, creating a dynamic interplay between transport processes and network growth.

Significant progress has been made in studying the structure and optimization of transport networks \cite{bernot_optimal_2009}, as well as the relation of the remodeling rules to cost function optimization \cite{chang2017}.
Most notably, Hu and Cai \cite{hu_adaptation_2013} proposed an ODEs model of a flow network represented as a discrete graph
, where the conductivity of each edge dynamically adapts based on local flow.
They demostrated that this adaptation rule minimizes an energy functional first proposed by Murray \cite{murray_physiological_nodate}, constrained by Kirchhoff law, incorporating the work required to drive the flow and the metabolic cost of maintaining the network.
Hu and Cai also proposed a corresponding PDE-based model of adapting transport networks represented as a conductivity field on a spatial continuum, which was subsequently studied by Haskovec, Markovich, and others (see \cite{haskovec_mathematical_2015, albi_biological_2016, haskovec_notes_2016, burger_mesoscopic_2019, haskovec_emergence_2023, astuto_self-regulated_2024, astuto_asymmetry_2025} and references therein).
Theoretical connections between both models have also been rigorously established in \cite{bellomo_active_2017, haskovec_ode-_2019, haskovec_rigorous_2019}.

The local adaptation and optimization regime proposed by Hu and Cai has also been extended by several authors.
Ronellenfitsch and Katifori \cite{ronellenfitsch_global_2016} introduced expanding edge lengths, demonstrating how network growth can further optimize transport efficiency and better explain leaf venation and animal vascular patterns.
Kramer and Modes \cite{kramer_biological_2023} incorporate metabolite demand and delivery,
highlighting that overly optimized flow networks may inadequately support metabolite delivery.

Despite these advances, existing models fail to account for the mass that must be delivered to growing vessels, as well as the material that may be released back into the system when vessels “self-destruct” during remodeling.
For example, it would be reasonable to assume that the released material is more readily available near the site of vessel removal, rather than being uniformly redistributed to distal locations.
A related phenomenon can be relevant for river networks and other branching systems where erosion and sediment deposition is spatially localized \cite{church2015morphodynamics}.

To uncover the fundamental principles governing the growth of biological transport networks, it is crucial to understand the interaction between local transport dynamics and the evolving network structure.
To the best of our knowledge, one such model exists for a simple one-dimensional (1D) domain without branches, as described in \cite{Graham2006}.
This model examines the outgrowth dynamics of a neurite, driven by the tip concentration of tubulin---the monomeric component of microtubules, which provide structural support and facilitate intracellular transport by molecular motors.
However this model does not account for branching and the accompanying the intricacies of a complex tree topology, 
including the possibility of branch retraction.
Inspired by this, we aim to develop an extended model that applies broadly to transport networks that utilize their own functionality to grow.

In this work, we describe a 1D transport network model in which tip growth is explicitly dependent on the local concentration of transported material.
To explore the utility of this model for investigating the interplay between transport and growth, we propose a numerical simulation framework and provide an illustrative example: the retraction of a branch to enhance the growth of another.
However, numerical observations suggest that a simple tip growth rules may not suffice to confer any growth advantage through branch retraction.
We further demonstrate that this limitation can be overcome by coupling the tip growth dynamics with an oscillating network.

\begin{figure}
	\includegraphics[width=\linewidth]{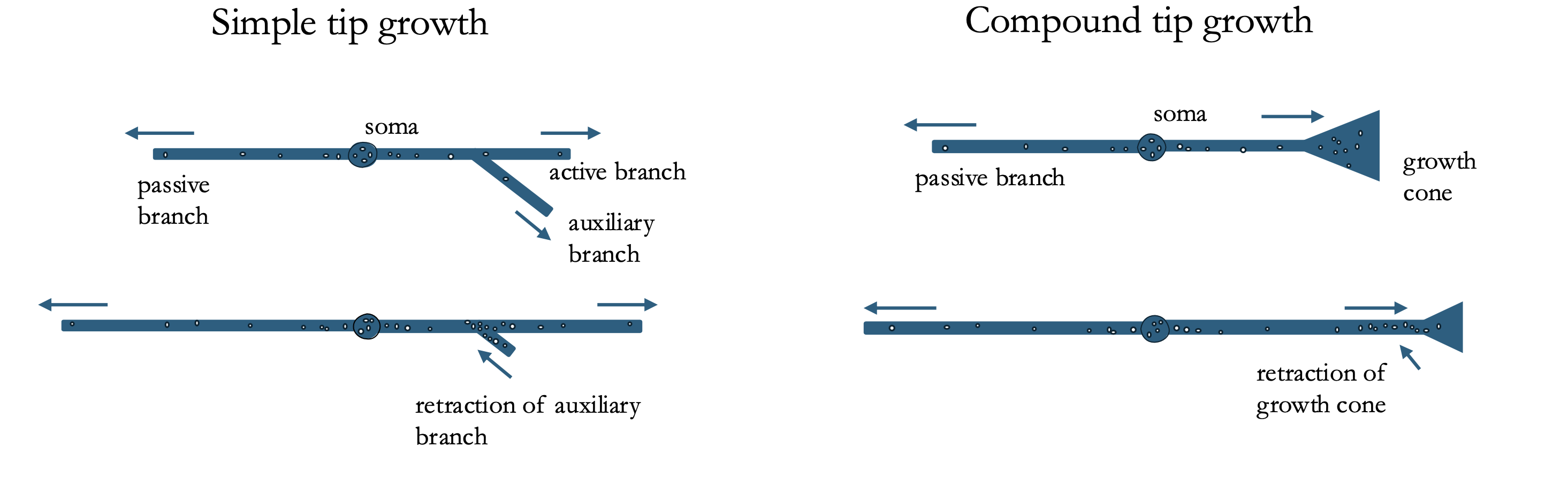}
	\caption{Schematic of Simple (left) and Compound (right) tip  growth construction and destruction process.}\label{fig:introfigure}
\end{figure}

\section{1D network model}

\subsection{General framework}

\begin{figure}
	\includegraphics[width=0.7\linewidth]{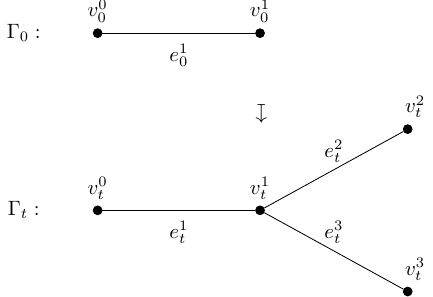}
	\caption{An example of an evolving metric graph $\Gamma_t$. Note that $|\mathcal{V}_t|$ and $|\mathcal{E}_t|$ can decreases or increases with respect to $t$ and vertices (resp. edges) with the same index represent analogous loci.}\label{fig:metric_graphs}
\end{figure}

We consider the time evolution of a transport network represented as a collection of metric graphs $\{\Gamma_t\}_{t\geq 0}$ indexed by time $t$ (see Figure~\ref{fig:metric_graphs}).
A metric graph $\Gamma_t$ is a collection of
the edges $\mathcal{E}_t=\{e_t^i\}$ identified with line segments with Euclidean metric,
and the vertices $\mathcal{V}_t=\{v_t^i\}$ consisting of equivalence classes of the boundary of all the edges $\bigcup_i\partial e_t^i$.
Here, we use the superscript notation for each edge and vertex to indicate the correspondence across the time, i.e., $e^i_t$ corresponds to $e^j_{t'}$ if $i=j$ and vice versa.
In this model, we will only consider a metric graph of which each edge is not a loop; otherwise, we can replace the loop by a cyclic chain of $2$ or more vertices.
For each edge $e^i_t$, we can specify a bond (directed edge) $b_t^i$ and its inverse denoted by $\bar{b}_t^i$ to give directionality to the graph, and we write $\mathcal{B}_t$ the set of all bonds of $\Gamma_t$ and their inverses.
Note that the choice of directionality only matters when there is an advection on the metric graphs.
We will denote by $x^b\in (0,L_e)$ the local coordinate of the bond $b$ of an edge $e\in \mathcal{E}_t$ with edge length $L_e=L_b=L_{\bar{b}}$, and its anti-coordinate by $x^{\bar{b}}$ so that the corresponding location on both bonds are given by the map $\phi$ on $\bigcup_t \mathcal{B}_t$ with $\phi|_b: b\to \bar{b}, x^b \mapsto L_e-x^{\bar{b}}$.
We say that $f:\mathcal{B}_t \to X$ is a ($X$-valued) function on $\Gamma_t$, abbreviated as $f:\Gamma_t \to X$, if $f|_b= f\circ \phi|_b$ for all $b\in \mathcal{B}_t$.
As such, it makes sense to view $f$ on the edges without mentioning the bonds.

We consider free material of concentration $c(\cdot,t):\Gamma_t\to\mathbb{R}_{\geq 0}$ satisfying advection-diffusion dynamics:
\begin{equation}
	\partial_t c + \partial_x (a c - D \partial_x c) = s \quad \text{in}\quad \Gamma_t.
	\label{eq:interior}
\end{equation}
This notation is to be understood as follows.
Here, $D(\cdot,t):\Gamma_t\to \mathbb{R}_+$ is the diffusion coefficient, $s(\cdot,t):\Gamma_t\to \mathbb{R}$ is a source (or sink) term, and the advection rate $a(\cdot,t):\mathcal{B}_t\to \mathbb{R}$ satisfies $a = - a \circ \phi$.
The equation \eqref{eq:interior} is to be interpreted in the sense of Euclidean 1D space on each bond, i.e.,
\begin{equation*}
	\partial_t c(x^b,t) + \partial_{x^b} (a(x^b,t) c(x^b,t) - D(x^b,t) \partial_{x^b} c(x^b,t)) = s(x^b,t),
\end{equation*}
with imposed \textit{vertex conditions}:
\begin{enumerate}
	\item $c(\cdot,t)$ is continuous in $\Gamma_t$---particularly, at all vertices,
	\item the fluxes $J_b := a c - D\partial_{x^b} c$ satisfy at each vertex $v\in \mathcal{V}_t$: \begin{equation}\label{eq:flux-VC}
		      \sum_{b\in \mathcal{B}^v} J_b(v,t)
		      = F_v[c],
	      \end{equation} 
	      where $\mathcal{B}^{v}$ denotes the outgoing bonds associated with $v$, and $F_v$ is a given functional of the concentration.
\end{enumerate}
The simplest case of $F_v$ would be $F_v \equiv 0$ for all $v$,
which corresponds to the so-called Neumann-Kirchoff condition which conserve the total material being transported.
More generally, when $F_v=\alpha c(v)$ for $\alpha\in \mathbb{R}\cup \{\infty\}$, this is known as $\delta$-type vertex condition is well-studied in the quantum graph theory \cite{berkolaiko_introduction_2013}.
Note that the way we describe \eqref{eq:flux-VC} is so that $F_v$ could depend non-locally on the concentration which allows us to compartmentalize the transport network
and couple the connected compartments back together in a particular way,
as we shall see in the subsequent section \S\ref{sec:growth_at_junction}.

\subsection{Tip growth rule}

Next, we describe a simple \textit{tip growth rule}.
We will only specify the growth condition on a subset of the degree $1$ vertices and their associated edges; the lengths of the remaining edges are fixed.
Such vertices will be refered to as the \textit{tips}, denoted by $\mathcal{T}_t \subset \mathcal{V}_t$.
For $b$ of which the terminal vertex, denoted by $\mathrm{t}(b)$, belongs to $\mathcal{T}_t$,
the length dynamics is governed by
\begin{equation}\label{eq:tipgrowth}
	\dot{L}_b = \gamma_v c(v,t) - \rho_v, \quad v=\mathrm{t}(b)\in \mathcal{T}_t,
\end{equation}
where we define
\begin{equation}
	\dot{L}_{b_t^i} := \lim_{h\to 0} \frac{L_{b_{t+h}^i}-L_{b_t^i}}{h}.
\end{equation}
The parameters defining the tip growth rule are the growth rate $\gamma_v\geq 0$ and the retraction rate $\rho_v\in\mathbb{R}$.
Taking into account that the growth only occurs at the tips---not anywhere in the interior of the edge, and supposing that the construction cost rate is $\kappa>0$, we have for $v=\mathrm{t}(b)\in \mathcal{T}_t$:
\begin{equation}
	\begin{split}
		F_v[c] &= -\left(c(v,t) + \kappa\right)\dot{L}_b(t)\\
		&=-\gamma_v c(v,t)^2-(\kappa-\rho_v)c(v,t)+\kappa\rho_v.
	\end{split}
	\label{eq:bdry}
\end{equation}
The reader may notice that when a branch is retracting, i.e., $\dot{L}_b<0$, material is returned to the system at the same conversion rate as the construction.
That is, we allow recycling of the material.
We will henceforth refer to $c$ as the \textit{free} material and $\kappa L_e$ the \textit{converted} material in an edge $e$.

We also require that a tip $v^i_t\in \mathcal{T}_t$ exists only if the length $L_{b}$ is positive for all $s\leq t$ for which $\mathrm{t}(b)=v^i_s\in \mathcal{T}_s$, so that the condition \eqref{eq:tipgrowth} makes sense.
In particular, if $t_*$ is the first time at which $L_{v^i_t}\searrow 0$ as $t\nearrow t_*$ while the other edge lengths remain positive, then $v_t^i$ disappears at $t=t^*$ at which $|\mathcal{V}_t|$ reduces by $1$.
This is the \textit{vanishing length condition}.
Moreover, depending on the problem, one may impose a condition which introduces an additional vertex (together with an additional edge) to $\mathcal{T}_t$.
For example, we could assume that $\{\mathcal{T}_t\}_{t\geq 0}$ is a continuous time binary branching process with certain birth rates with death according to the vanishing length condition.

Finally, we propose a finite difference method to numerically solve this problem and is applied in the next section.
This can be found in Appendix~\ref{sec:appendix}.

\section{Retraction-induced growth enhancement}

\subsection{Simple tip growth}

With a general modeling framework in place for exploring the interaction between network shape and transport dynamics, we now examine a specific case to assess its utility.
We consider one intuitive idea that, within the same transport network, a branch with multiple growing tips may have a growth advantage over a branch with a single tip.
The heuristics for this is that more material would be delivered to a larger expanding front with the help of diffusion, as opposed to a system with pure advection.
Thus, a network structure with multiple tips could store more (free and converted) material,
potentially facilitating greater overall elongation.

\begin{figure}
	\includegraphics[width=\linewidth]{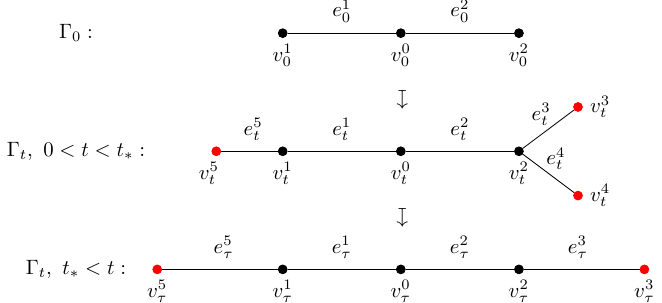}
	\caption{
		Setup for the simple tip growth.
		Here, the tips are labeled red.
		At time $0<t<T$, the two tips supported on the right-hand side grows at the same rate until time $t_*\in (0,T)$, after which $v^4_t$ retracts until it reaches $v^2_t$.
	}
	\label{fig:setup}
\end{figure}

To test this hypothesis, we set up a local tip growth problem and implement a numerical experiment.
Consider an initial network $\Gamma_0=(\mathcal{V}_0,\mathcal{E}_0)$ consisting of two edges $e_0^1, e_0^2$ of the same length $L_{e_0^1}=L_{e_0^2}$ extending from a common vertex $v^0_0$ to $v^1_0$ and $v^2_0$ respectively (Figure~\ref{fig:setup}).
All free material is initially concentrated at $v_0^0$, i.e.,
\begin{subequations}\label{eq:init_tipgrowth}
	\begin{equation}
		c(x,0)\,dx=C_0\delta_{v^0_0}
	\end{equation}
	where $C_0>0$ is the total initial amount of the free material and $\delta_{v^0_t}$ denotes the Dirac measure on $\Gamma_t$ at $v^0_t$.
	We are comparing the growth on both sides of the the initial source at $v_t^0$.
	The key difference between the two is that $e_t^1$ supports the growth of two tips forming a Y-shaped branch, while $e^2_t$ has only a single tip.
	More precisely, for $t$ near $0$, we have two more edges $e_t^3, e_t^4$ connecting $v_t^2$ to $v_t^3,v_t^4$, and $e_t^5$ connecting $v_t^1$ to $v_t^5$ respectively.
	The tips during this time is $\mathcal{T}_t=\{v_t^3, v_t^4,v_t^5\}$, for which we have
	\begin{equation}
		(L_{v_t^3}, L_{v_t^4},L_{v_t^5}) \to (0,0,0) \quad\text{as}\quad t\to 0.
	\end{equation}
\end{subequations}
The free material is then distributed between the two branches via an advection-diffusion process \eqref{eq:advdiff} with constant coefficients:
\begin{subequations}\label{eq:param_tipgrowth}
	\begin{equation}
		a(x^b,t)\equiv a \geq 0,\quad D(x^b,t)\equiv D>0,\quad\text{and}\quad s(x^b,t)\equiv 0
	\end{equation}
	for all $t>0$ and any bond $b$ directed away from $v^0_t$.
	The tips $\mathcal{T}_t$ satisfies the tip growth rule \eqref{eq:tipgrowth}, with constant $\gamma_v\equiv\gamma_0>0$ and $\rho_v\equiv 0$ and its corresponding vertex conditions \eqref{eq:bdry}, and the non-tip vertices satisfy the Neumann-Kirchoff condition, i.e., $F_v\equiv 0$.
	We expect that the Y-shaped branch ($e^2_t, e^3_t, e^4_t$)
	will retain more material due to the increased number of paths for diffusion.
	After some time $t_*$, one of the expanding edges---say $e^4_t$, $t>t_*$---retracts, and the remaining tip $v_t^3$ should be able to utilize the stored material, potentially gaining a growth advantage for $e^3_t$ comparing to $e^2_t$ on the other side.
	For this, we let
	\begin{equation}
		(\gamma_{v^3_t},\gamma_{v^4_t},\gamma_{v^5_t}) = \begin{cases}
			(\gamma_0,\gamma_0,\gamma_0), & \text{ if }t<t_*,                         \\
			(\gamma_0,0,\gamma_0),        & \text{ if }t_*<t\text{ and } L_{e^4_t}>0, \\
		\end{cases}
	\end{equation}
	\begin{equation}
		(\rho_{v^3_t},\rho_{v^4_t},\rho_{v^5_t}) = \begin{cases}
			(0,0,0),      & \text{ if }t<t_*,                         \\
			(0,\rho_0,0), & \text{ if }t_*<t\text{ and } L_{e^4_t}>0. \\
		\end{cases}
	\end{equation}
	where $\rho_0>0$.
\end{subequations}

To summarize, the problem reads
\begin{subequations}\label{eq:simple_tip_growth_summary}
	\begin{equation}
		\partial_t c +  a \partial_x c - D \partial_{xx} c = 0 \quad \text{in}\quad \Gamma_t,\quad t>0,
	\end{equation}
	\begin{equation}
		\dot{L}_{e^i_t} = \begin{cases}
			\gamma_{v^i_t} c(v^i_t,t) - \rho_{v^i_t}, & \text{if}\quad i=3,4,5, \\
			0,                                        & \text{otherwise}
		\end{cases}
	\end{equation}
	\begin{equation}
		\sum_{b\in \mathcal{B}_{v^i_t}} J_b(v^i_t,t) = \begin{cases}
			-\left(c(v^i_t,t) + \kappa\right)\dot{L}_{e^i_t} & \text{if}\quad i=3,4,5, \\
			0                                                & \text{otherwise},
		\end{cases}
	\end{equation}
\end{subequations}
with initial conditions in \eqref{eq:init_tipgrowth} and parameters in \eqref{eq:param_tipgrowth}.

\begin{figure}
	\begin{center}
		\includegraphics[width=0.95\textwidth]{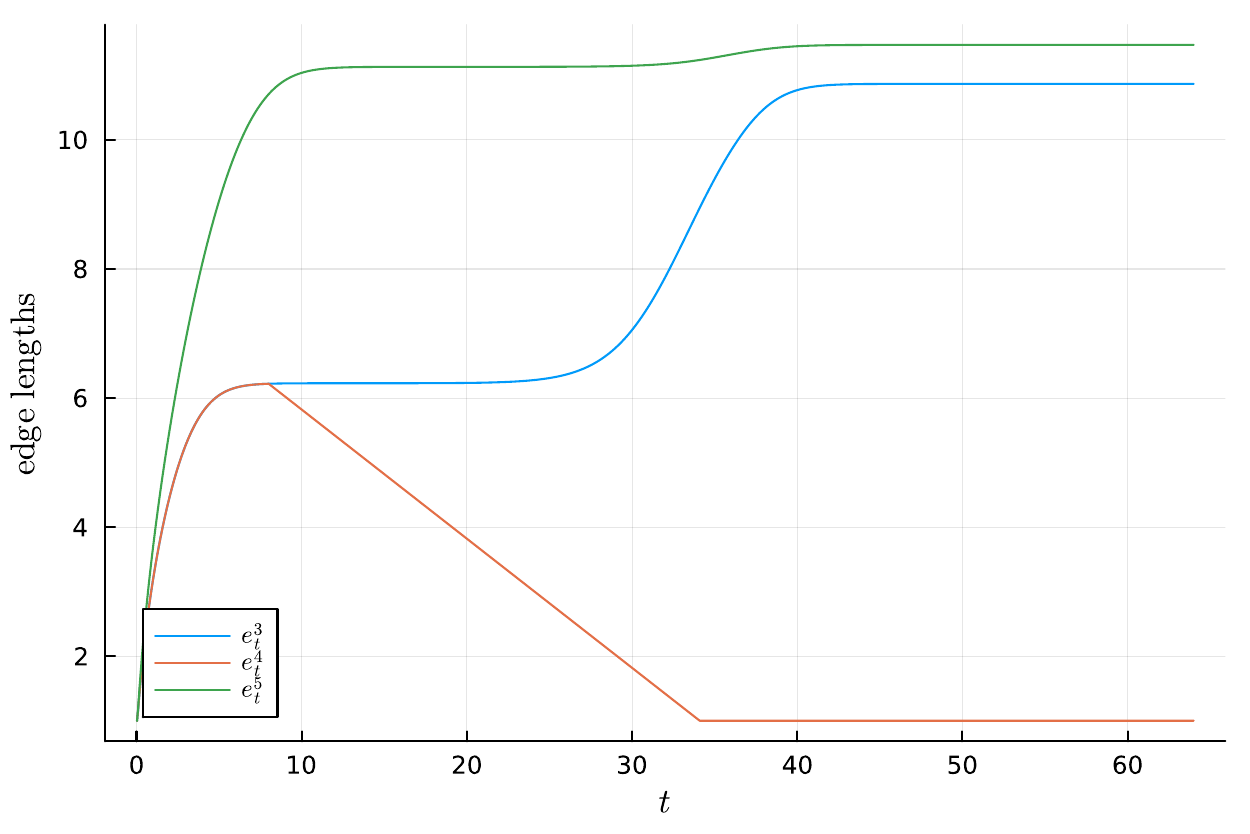}
	\end{center}
	\caption{The length from the origin to each growing tip satisfying \eqref{eq:simple_tip_growth_summary}.}\label{fig:snapshots}
\end{figure}

\begin{figure}
	\begin{center}
		\includegraphics[width=0.95\textwidth]{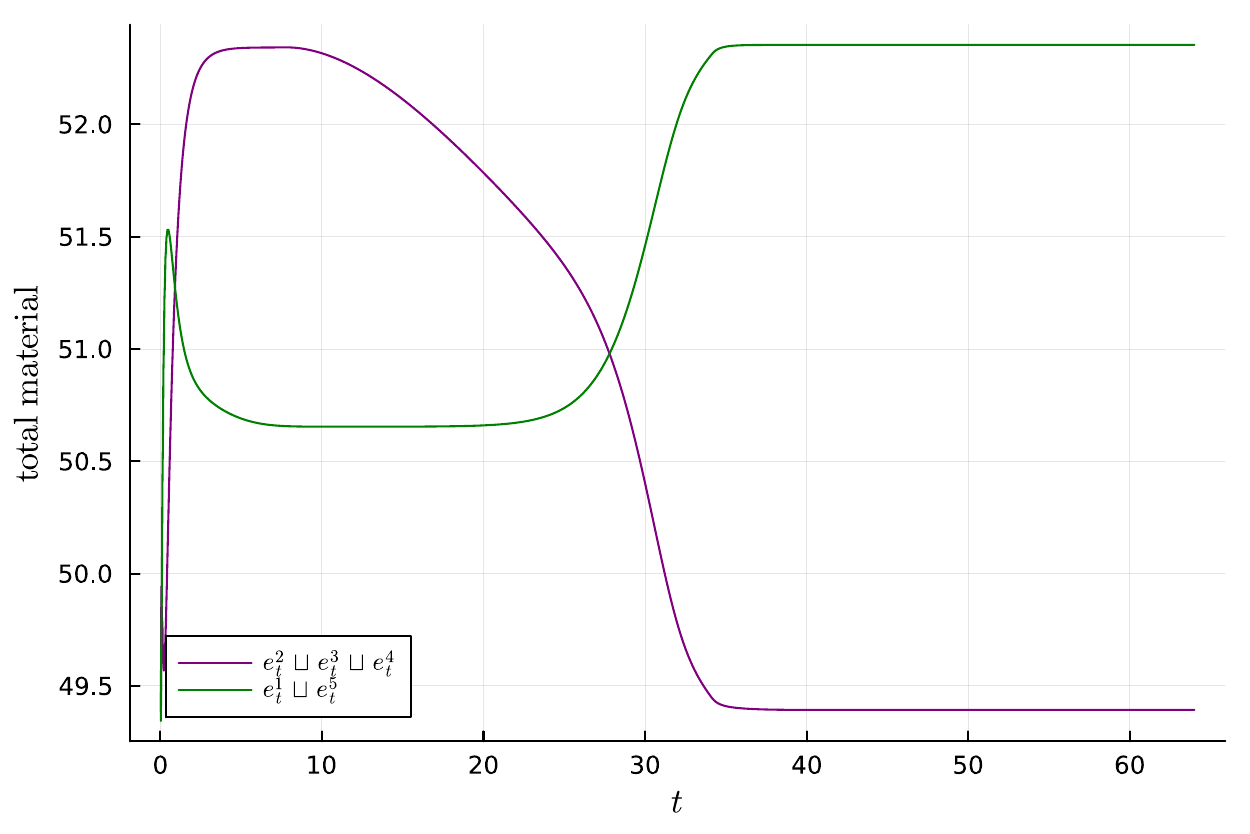}
	\end{center}
	\caption{The dynamics of the total material, including both free and converted material, in each branch supported at $v^0_t$.}\label{fig:material-distribution}
\end{figure}

We numerically solved the system using the finite difference method described in Appendix~\ref{sec:appendix}.
The caveat is that it is impossible to simulate at the time when the length of an edge is too small.
Thus, to make the problem subject to our numerical method we include the edges $e^3_t,e^4_t,e^5_t$ in $\Gamma_t$ for all $t\geq 0$, and we let the initial length
\begin{equation}
	L_{e^3_0}=L_{e^4_0}=L_{e^5_0}=\varepsilon 
\end{equation}
and
\begin{equation}
	\rho_{v^4_t} = 0 \quad\text{if}\quad L_{e^4_t} < \varepsilon.
\end{equation}
for small truncation parameter $\varepsilon>0$.
Initially, before the retraction, the Y-shaped branch is shorter due to simultaneous growth of $2$ tips (Figure~\ref{fig:snapshots}), but it retains more material as predicted (Figure~\ref{fig:material-distribution}).
However, contrary to our intuition, $L_{e^3_t}$ remains shorter than $L_{e^5_t}$ even after retraction.
As shown in Figure~\ref{fig:material-distribution}, this observation coincides with the escape of the stored material from the retracting edge $e^4_t$ toward the single-tip branch $e^5_t$,
preventing $e^3_t$ from overcoming its initial length disadvantage.


\begin{figure}
	\begin{center}
		\includegraphics[width=0.95\textwidth]{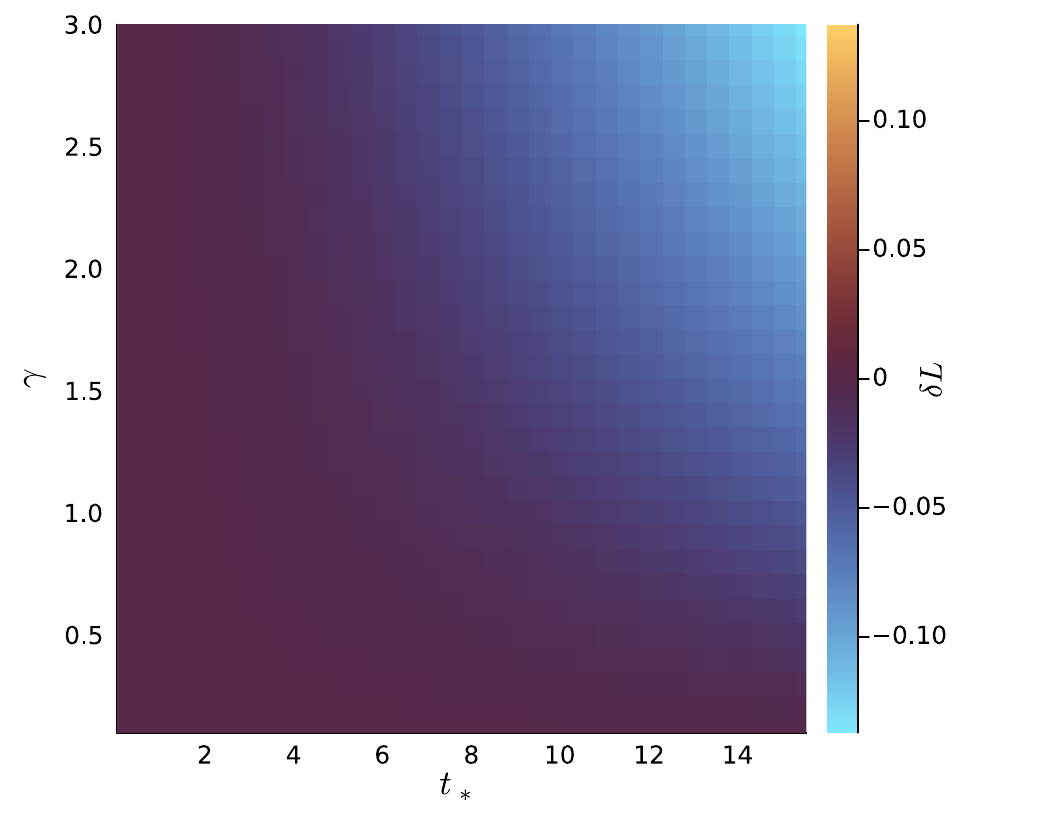}
	\end{center}
	\caption{Failure of retraction-induced growth enhancement in the problem \eqref{eq:simple_tip_growth_summary}.
	Here, the length difference $\delta L=L_{e^3_T}-L_{e^5_T}$ where $T=64$.
	}\label{fig:robustness}
\end{figure}

Interestingly, as shown in Figure~\ref{fig:robustness}, this result remains robust across different parameter choices, suggesting the existence of an underlying mathematical or physical principle that governs this limitation.
Understanding this mechanism warrants further investigation.

\subsection{Compound tip growth} \label{sec:growth_at_junction}

\begin{figure}
	\includegraphics[width=\linewidth]{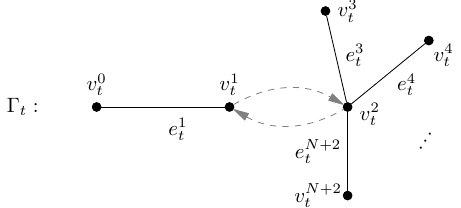}
	\caption{The structure of the compound tip. The oscillation at the growing tip is modeled as coupling between a branch with a simple tip and an oscillating $N$-star graph at $v^1_t$ and $v^2_t$ respectively.}\label{fig:coupled}
\end{figure}

The previous result suggests that we might need a mechanism which mitigates the escape of the material.
For this, we now consider a coupling between a growing tip and an oscillating star-shaped network (Figure~\ref{fig:coupled}).
This oscillating network is inspired by a growth cone---a structure on the tip of a growing neurite of a neuron consisting of small spike-like membrane protrusions called filopodia.
The oscillating network will periodically convert and store material, and directly feed it to the growing tip.
This circumvents the disadvantage in the previous set up that the material from the retracting edges has to be transported a long distance to the growing tip.
We shall see that this will indeed produce retraction-induced growth enhancement.

The network structure $\Gamma_t$, $t\geq 0$, now consists of two (topologically) connected components: an edge $e^1_t$ connecting a large pool of material at $v^0_t$ to a growing tip $v^1_t$, and an oscillating network which is an $N$-star graph centered at $v^2_t$ with tips $v^i_t$ for $i=3,\dots,2+N$ connected through $e^i_t$ (Figure~\ref{fig:coupled}).
Mathematically, the tips are $\mathcal{T}_t=\{v^1_t,v^3_t,\dots,v^{2+N}_t\}$, but from the physical point of view we could consider the oscillating network together with $v^1_t$ as parts of a ``compound'' growing tip of the branch $e^1_t$.
We choose the bonds $b^i_t$ with $i=1,3,...,2+N$ pointing away from $v^1_t$ or $v^2_t$ so that $\mathrm{t}(b^i_t)=v^i_t$.
The free material concentration $c$ satisfies \eqref{eq:advdiff} with the same parameter \eqref{eq:param_tipgrowth} as previously.
The lengths $L_{b^i_t}$ satisfy the same growth condition \eqref{eq:tipgrowth} with
\begin{subequations}\label{eq:param_compound}
	\begin{equation}
		\gamma_{v^i_t} = \begin{cases}
			\gamma_0,                                               & \text{ if }i=1,            \\
			\frac{\gamma_0}{2} \left(1+\sin(\frac{t}{\tau})\right), & \text{ if } i=3,\dots,N+2, \\
		\end{cases}
	\end{equation}
	\begin{equation}
		\rho_{v^i_t} = \begin{cases}
			0,                                                    & \text{ if }i=1,           \\
			\frac{\rho_0}{2} \left(1-\sin(\frac{t}{\tau})\right), & \text{ if } i=3,\dots,N+2
			\\
		\end{cases}
	\end{equation}
\end{subequations}
Here, for $v^i_t$, $i=3,\dots,N+2$, we impose an alternation between growth and retraction with a period $\tau$ and maximal rates $\gamma_0$ and $\rho_0$ respectively.

The vertex condition at the material source $v^0_t$ is given by
\begin{subequations}\label{eq:vc_compound}
	\begin{equation}
		F_{v^0_t}[c]= k_0(c_* - c(v^0_t,t))
	\end{equation}
	which can be interpreted as supplying the vertex $v_0^0$ with an infinite pool of material of cencentration $c_*$, and $k_0$ is the coupling strength.
	Since we are also coupling the growing tip $v^1_t$ with the oscillating network at $v^2_t$, we will modify the vertex condition \eqref{eq:bdry} by adding coupling terms to $F_{v^1_t}$ and $F_{v^2_t}$ from what would otherwise be the same local tip growth as \eqref{eq:bdry} and Neumann-Kirchoff conditions:
	\begin{equation}
		F_{v^1_t}[c] = -\left(c(v^1_t,t) + \kappa\right)\dot{L}_{b^1_t} - k_1 \left(c(v^1_t,t) - c(v^2_t,t)\right),
	\end{equation}
	\begin{equation}
		F_{v^2_t}[c] = k_1 \left(c(v^1_t,t) - c(v^2_t,t)\right),
	\end{equation}
	\begin{equation}
		F_{v^i_t}[c]=-\left(c(v^i_t,t) + \kappa\right)\dot{L}_{e^i_t}, \quad i=3,\dots,N+2.
	\end{equation}
\end{subequations}
Lastly, the tips of the oscillating network satisfy the same tip growth vertex condition \eqref{eq:bdry}.

In summary, the compound tip growth problem reads
\begin{subequations}\label{eq:compound_tip}
	\begin{equation}
		\partial_t c +  a \partial_x c - D \partial_{xx} c = 0 \quad \text{in}\quad \Gamma_t,\quad t>0,
	\end{equation}
	\begin{equation}
		\dot{L}_{e^i_t} =
		\gamma_{v^i_t} c(v^i_t,t) - \rho_{v^i_t},
	\end{equation}
	\begin{equation}
		\sum_{b\in \mathcal{B}_{v^i_t}} J_b(v^i_t,t) = F_{v^i_t}[c],
	\end{equation}
	with parameters in \eqref{eq:param_compound}, vertex conditions \eqref{eq:vc_compound}, and we consider the initial conditions:
	\begin{equation}
		c(\cdot,0)\equiv 0, \quad
		L_{e^1_0} = L_0, \quad
		L_{e^3_0}=\dots=L_{e^{N+2}_0} = L_1
	\end{equation}
	for some $L_1>L_0>0$.
\end{subequations}



\begin{figure}
	\begin{center}
		\includegraphics[width=0.95\textwidth]{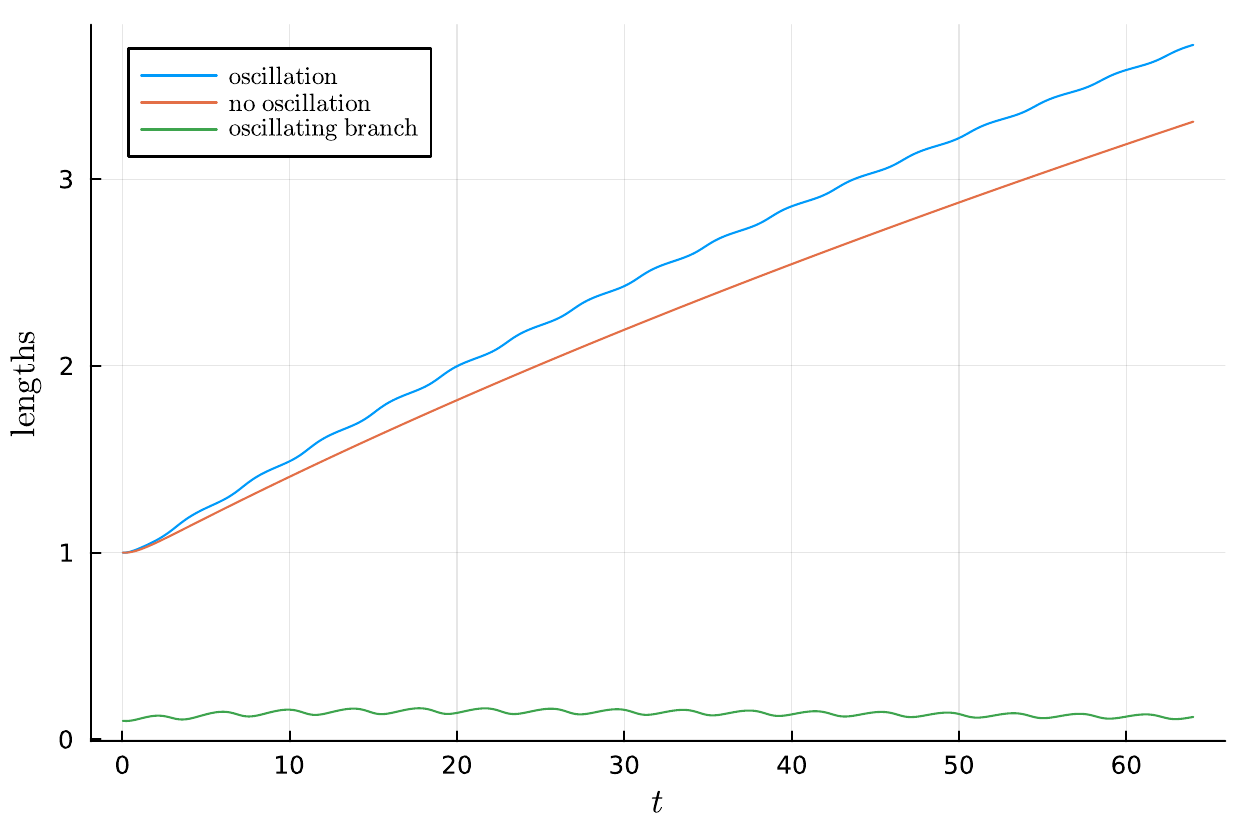}
	\end{center}
	\caption{The plot of Lengths of $L_{e^1_t}$ when there is coupling between the growing tip  and the oscillating network (blue, satisfying \eqref{eq:compound_tip}) and no coupling (orange, satisfying \eqref{eq:simple_tip}). The green line represents the length of one of the edges of the oscillating network.}\label{fig:coupled-full}
\end{figure}

We will compare a numerical solution of \eqref{eq:compound_tip} to the case when we do not have an oscillating network at the growing tip.
This is given by the following problem on $\Gamma_t=(\mathcal{V}_t,\mathcal{E}_t)$, where $\mathcal{V}_t=\{v^0_t,v^1_t\}$ and $\mathcal{E}_t = \{e^1_t\}$, choosing $b^1_t$ pointing away from $v^0_t$:
\begin{subequations}\label{eq:simple_tip}
	\begin{equation}
		\partial_t c +  a \partial_x c - D \partial_{xx} c = 0 \quad \text{in}\quad \Gamma_t,
	\end{equation}
	\begin{equation}
		\dot{L}_{e^1_t} =
		\gamma_0 c(v^1_t,t),
	\end{equation}
	\begin{equation}
		J_{b^1_t}(v^0_t,t) = k_0(c_*-c(v^0_t,t)),
	\end{equation}
	\begin{equation}
		J_{\bar{b}^1_t}(v^1_t,t) = -\left(c(v^1_t,t) + \kappa\right)\dot{L}_{e^1_t},
	\end{equation}
	\begin{equation}
		c(\cdot,0)\equiv 0, \quad
		L_{e^1_0} = L_0.
	\end{equation}
\end{subequations}
Using the implicit finite difference method in Appendix~\ref{sec:appendix} and parameters in the Table \ref{tab:param_compound},
\begin{table}
	\caption{Parameters scheme for the simple and compound tip growth}
	\begin{center}
		\begin{tabular}[c]{l|c|c}
			\hline
			\multicolumn{1}{c|}{\textbf{Parameters}} &
			\multicolumn{1}{c|}{\textbf{Simple tip}} &
			\multicolumn{1}{c}{\textbf{Compound tip}} \\
			\hline
			$\gamma_0$ & $0.6$ & $0.2$ \\
			$\rho_0$ & $0.2$ & $0.2$ \\
			$\kappa$ & $5.0$ & $5.0$ \\
			$k_0$ & N/A & $2.0$ \\
			$k_1$ & N/A & $1.0$ \\
			$D$ & $1.0$ & $5.0$ \\
			$a$ & $2.0$ & $0.0$ \\
			$c_*$ & N/A & $0.1$ \\
			$\tau$ & N/A & $\pi / 10$ \\
			$L_{e^1_0},L_{e^2_0}$ & $1.0$ & $1.0$ \\
			$L_{e^i_0}$, $i\geq 3$ & $\varepsilon=0.025$ & $0.1$ \\
			$N$ & N/A & $10$ \\
			$\varepsilon$ & $0.025$ & $0.025$       \\
			\hline
		\end{tabular}
	\end{center}
	\label{tab:param_compound}
\end{table}
we plot the lengths $L_{e^1_t}$ of both problems \eqref{eq:compound_tip} and \eqref{eq:simple_tip} in Figure \ref{fig:coupled-full}.
Indeed, we see that the growth of the compound tip exceeds that of the simple tip without oscillating network.

To explore the extent of this in the parameter space, we consider a reduction of the problem \eqref{eq:compound_tip}.

\subsubsection{Model reduction}
Under certain parametric scheme, namely when the diffusion timescale is much smaller than the oscillation timescale, we could reduce the system into a single ODE describing the length dynamics.
By assuming
\begin{equation*}
	\frac{L_0^2/D}{\tau} \ll 1,
\end{equation*}
we obtain the quasi-steady state advection-diffusion equation for the free material:
\begin{equation}\label{eq:advdiff}
	a\partial_x \tilde{c} - D\partial_{xx} \tilde{c} = 0\quad\text{in}\quad e^1_t.
\end{equation}
Moreover, we assume that the coupling strength $k_1$ is strong---in particular,
\begin{equation*}
	k_1\gg L_0D,
\end{equation*}
so that we approximate
\begin{equation}
	\tilde{c}(v^1_t,t) = \tilde{c}(v^2_t,t).
\end{equation}
Further, by assume in contrast that
\begin{equation*}
	\frac{\tau}{L_1^2/D}\ll 1,
\end{equation*}
we could approximate
\begin{equation}
	\partial_x\tilde{c} = 0 \quad\text{in}\quad e^i_t,\ i=3,\dots,N+2.
\end{equation}
Therefore, the concentration on the oscillating network is approximately constant:
\begin{equation}
	\tilde{c}|_{e^i_t} \equiv C(t),\quad i=3,\dots,N+2.
\end{equation}
Since the material fluxes on $b^1_t$ is constant, the vertex conditions \eqref{eq:vc_compound} for \eqref{eq:advdiff} at $v^0_t,v^1_t$ are reduced to
\begin{equation}\label{eq:bdry_gc}
	J_{b^1_t}:=a\tilde{c} - D\partial_x \tilde{c} \equiv  
	k (c_*-\tilde{c}(0,t)) 
	= (C(t)+\kappa) \left(\dot{L}_{e^1_t}+\sum_{i=3}^{N+2}\dot{L}_{e^i_t}\right).
\end{equation}
For simplicity, we modify the length dynamics so that
\begin{equation}\label{eq:reduced_growth}
	\dot{L}_{e^1_t} = \gamma C,\quad \dot{L}_{e^i_t}=\gamma C \sin (t/\tau),\quad i=3,\dots,N+2,
\end{equation}
Here, in \eqref{eq:bdry_gc} and \eqref{eq:reduced_growth}, we remove the $0$ subscription from $k_0$ and $\gamma_0$ to avoid cluttering of notations.


We will express $C$ in terms of $L_{e^1_t}$ so that the model is reduced to a single ODE.
Suppose for now that $a\neq 0$.
For notational convenience, we simply write $L=L_{e^1_t}$.
The general solution to the advection-diffusion \eqref{eq:advdiff} reads
\begin{equation}\label{eq:shaftconc}
	\tilde{c}(x,t)= \tilde{c}(0,t) + \partial_x \tilde{c}(0,t)\frac{D}{a} \left(\exp \left(\frac{ax}{D}\right)-1\right).
\end{equation}
For $x=0$, the boundary condition \eqref{eq:bdry_gc} gives
\begin{equation}\label{eq:rootconc}
	\tilde{c}(0,t) = \frac{D}{a+k} \partial_x \tilde{c}(0,t) + \frac{kc_*}{a+k}.
\end{equation}
Substituting \eqref{eq:rootconc} into \eqref{eq:shaftconc} and using the last equality of \eqref{eq:bdry_gc}, we get
\begin{equation}\label{eq:slope}
	\partial_x \tilde{c}(0,t) = \frac{(a+k)C(t)- kc_*}{D + \frac{(a+k)D}{a} \left(\exp \left(\frac{La}{D}\right) - 1\right)}
\end{equation}
Using \eqref{eq:slope} and \eqref{eq:rootconc}, we arrive at
\begin{equation}
	C(t) = \frac{\frac{ak}{a+k}c_* \left(1+ \frac{a+k}{a} \left(\exp \left(\frac{La}{D}\right)-1\right)\right)+ \frac{k^2c_*}{a+k}}{\kappa\gamma \left(1 + N  \sin \frac{t}{\tau}\right) \left(1+ \frac{a+k}{a} \left(\exp \left(\frac{La}{D}\right)-1\right)\right)+k}
\end{equation}
Similarly, for $a=0$, we have
\begin{equation}
	C(t) = \frac{kc_*}{\kappa\gamma\left(1 + N  \sin \frac{t}{\tau}\right) \left(1+ \frac{kL}{D}\right) + k}
\end{equation}
Thus, we obtain
\begin{equation}\label{eq:lengthODE}
	\dot{L} =
	\begin{cases}
		\frac{\frac{ak}{a+k}c_* \left(1+ \frac{a+k}{a} \left(\exp \left(\frac{La}{D}\right)-1\right)\right)+ \frac{k^2c_*}{a+k}}{\kappa \left(1 + N  \sin \frac{t}{\tau}\right) \left(1+ \frac{a+k}{a} \left(\exp \left(\frac{La}{D}\right)-1\right)\right)+k\gamma} & \text{for}\quad a \neq 0, \\
		\frac{kc_*}{\kappa\left(1 + N  \sin \frac{t}{\tau}\right) \left(1+ \frac{kL}{D}\right) + k\gamma}                                                                                                                                                            & \text{for}\quad a = 0.
	\end{cases}
\end{equation}

\begin{figure}
	\begin{center}
		\includegraphics[width=0.95\textwidth]{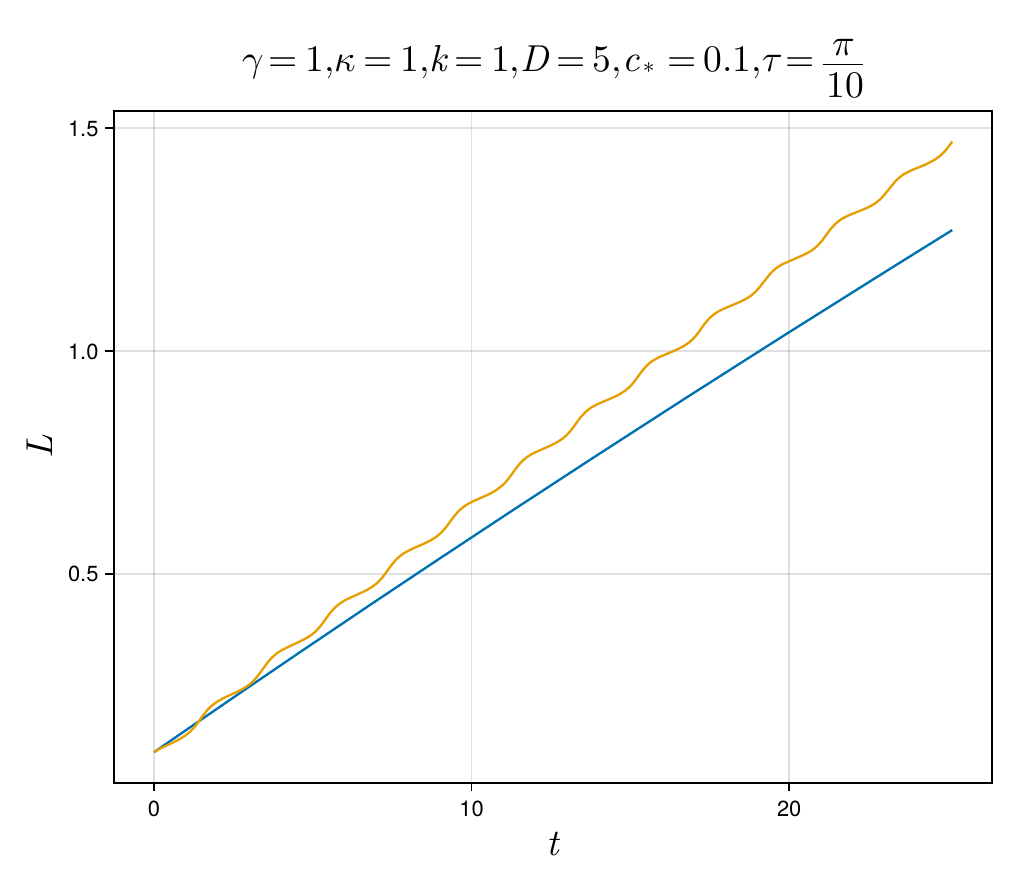}
	\end{center}
	\caption{Comparison between the lengths $L$ satisfying \eqref{eq:lengthODE} when $N=0$ and $N=1$.}\label{fig:RIGE}
\end{figure}

Using implicit Euler's method and parameters in the Table \ref{tab:param},
\begin{table}
	\caption{Parameters scheme for RIGE}\label{tab:param}
	\begin{center}
		\begin{tabular}[c]{l|l}
			\hline
			\multicolumn{1}{c|}{\textbf{Parameters}} &
			\multicolumn{1}{c}{\textbf{Values}}                   \\
			\hline
			$\gamma$                                 & $1$        \\
			$\kappa$                                 & $1$        \\
			$k$                                      & $1$        \\
			$D$                                      & $5$        \\
			$c_*$                                    & $0.1$      \\
			$\tau$                                   & $\pi / 10$ \\
			$L(0)$                                   & $0.1$      \\
			\hline
		\end{tabular}
	\end{center}
\end{table}
we plot the solutions of when $N=0$ and $N=1$ in Figure \ref{fig:RIGE}.
We see that this indeed does capture the previous observation that having an oscillating network at on the compound tip facilitate the growth.
To demonstrate the robustness of the compound tip growth, we plot the length difference between a compound tip and a simple tip at time $T$ over the $(\tau,\gamma)$-plane.
We see that the as $\tau$ or $\gamma$ approches $0$, the effect of retraction-induced growth enhancement diminishes.
On the other hand, the length difference seems to depend monotonously on $\gamma$ while we expect that the length difference should vanish as $\tau$ increases.

Unlike in the case of simple tip growth, 
when the retraction happens, the withdrawn material is delivered exactly at the place where it is needed without having to travel further to reach the growing tip.
In other words, this model circumvent the length disadvantage which piles up as the branch elongates.
This gives us an insight on how biological network could leverage retraction of branches in favor of growth.

\begin{figure}
	\begin{center}
		\includegraphics[width=0.95\textwidth]{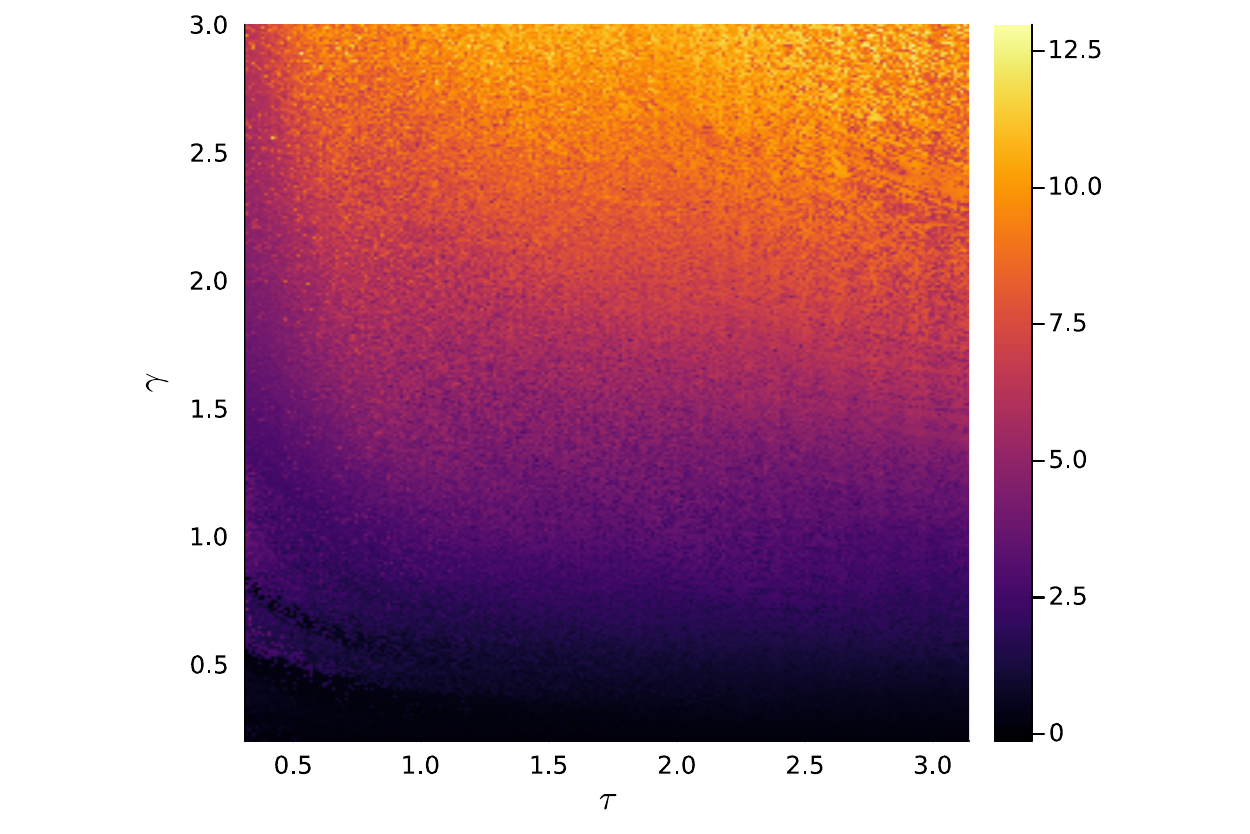}
	\end{center}
	\caption{The plot of the length difference $\delta L= \ell_1-\ell_0$ of $L=\ell_1$ satisfying \eqref{eq:lengthODE} when $N=1$, and $L=\ell_0$ when $N=0$.}\label{fig:compound-phase-plane}
\end{figure}

\section{Discussion}

In this work, we have introduced a transport network model with a growth rule applied at vertices.
The model describes how material concentration at the tips influences the elongation of branches, leading to emergent network structures.
By formulating the transport dynamics through an advection-diffusion process and implementing a vertex-based growth mechanism, we have explored how the interplay between transport and local growth dynamics shapes the network evolution.

A central result of this study is the observed failure of retraction-induced growth enhancement in the case of fixed branching points.
While an oscillating tip can facilitate growth enhancement by redistributing material effectively, a fixed branching point does not exhibit this behavior.
This suggests that for growth enhancement via retraction to occur, the network must possess a mechanism to prevent excessive material escape from the retracting branch.
The persistence of this excessive escape remains an open question, warranting further analytical and numerical investigation.
Moreover, the robustness of this phenomenon across different parameter choices indicates that it is not merely an artifact of specific model settings but may instead be governed by a deeper mathematical or physical principle.

Several simplifying assumptions in our model introduce potential limitations.
First, the branching rule employed is ad hoc, meaning that the formation of new tips is not determined dynamically but rather imposed externally.
A more natural extension would be to develop a rule in which branching arises organically from the transport and growth dynamics.
Additionally, our current model assumes that advection is a secondary effect, whereas in biological systems such as vascular networks, advective transport often plays a dominant role in material redistribution.
Incorporating stronger advection effects may lead to qualitatively different growth behaviors.

Another key assumption is that production occurs only at specific locations, such as at tips or a central reservoir. In real biological networks, production may be distributed throughout the system. Investigating how decentralized production influences growth patterns could provide deeper insights into self-organized network formation. Moreover, growth itself can induce mass flow, as suggested by prior work on advection-driven transport in growing networks \cite{heaton_advection_2012}. Integrating such effects may yield a more comprehensive understanding of how biological networks develop.

Beyond growth models that are dominated by the effects of transport, other biological networks exhibit alternative growth mechanisms. For example, the lymphatic vessel nertwork of the mouse ear achieves its final form via VEGF
dependent side-branching from existing capillaries by directionally targeting low-density regions \cite{Ucar2023}, indicating that material transport in the system at this stage of development plays at most a secondary role. 
Incorporating a broader class of local growth rules into our model would allow us to compare and contrast different strategies for network expansion and adaptation.

Additionally, scaling laws play a crucial role in biological transport networks.
A natural extension of this work would be to explore how scaling principles could be incorporated into the advection rate and diffusion coefficient, potentially leading to self-similar or optimized network structures.
%

A particularly intriguing open question is whether there exists a growth rule that results in a network optimized for a given cost function.
Many biological systems evolve to balance efficiency, robustness, and resource expenditure, suggesting that certain growth mechanisms may inherently favor optimal configurations.
Identifying such principles within our framework could provide insights into the broader class of adaptive transport networks observed in nature.

More generally, this study highlights the intricate coupling between transport and growth in dynamic networks.
Understanding these interactions not only sheds light on biological development but may also inspire new approaches in artificial and engineered systems, such as adaptive material design, vascularized tissue engineering, and self-organizing infrastructure networks.

\section*{Acknowledgement}
CS, EK and YM were supported by the MRSEC grant MRSEC-DMR2309043 (USA), and CS and YM by the Math+X award from the Simons Foundation (Award ID 234606). EK acknowledges support from the HFSP Award 977405.
\printbibliography

\appendix%

\section{Numerical method}\label{sec:appendix}
\AB{Comment on when lengths are too short}

The main challenge in finding numerical solution to our problem is that the structure of the graph can change dynamically.
To the best of our knowledge, there is only numerical study of static metric graph \cite{arioli_finite_2018}.
We thereby propose a finite difference method for such a problem.

We consider metric graphs $\{\Gamma_t\}_{t_0<t<t_*}$ which are mutually homeomorphic so that we have a homeomorphism $\Gamma_t \to \Gamma_*$ for each $t$, for some metric graph $\Gamma_*$.
Particularly, $(t_0,t_*)$ is a time period between branching events or complete retraction of a branch.
We will shorten the notation $L_{e^i_t}$ as $L_t^i$ and $x^{b^i_t}$ by $x^i_t\in (0,L_t^i)$.
On the other hand, the local coordinate $\xi^i\in(0,1)$ of the corresponding bond $b^i_*$ of $\Gamma_*$ is time independent.
Then, the change of local coordinate maps given by $x^i_t\mapsto x^i_t/L^i_t$ define such homeomorphisms.
With this we can rewrite the equation \eqref{eq:interior}:
\begin{equation}
	\partial_t \hat{c} + \frac{1}{L}\partial_\xi \left( \left(\hat{a} - \xi \dot{L}\right)\hat{c} - \frac{\hat{D}}{L}\partial_\xi \hat{c}\right) + \frac{\dot{L}}{L}\hat{c} = \hat{s}\quad\text{in}\quad \Gamma_*.
	\label{eq:intstar}
\end{equation}
Here, 
$L(\cdot,t):\bigcup\mathcal{B}_*\to \mathbb{R}_+, \xi^i\mapsto L^i_t$ is the length of the edges and vice versa for $\dot{L}$;
the $\hat{\cdot}$ notation transfers data from $\Gamma_t$ to $\Gamma_*$, 
i.e., 
$u(x^i_t,t) = \hat{u}(\xi^i,t)$ for $u\in \{c,a,D\}$. 
Since the problem is not well-defined at time to which $L\to 0$, possibly at $t_0$ or $t_*$, we can only implement the numerical method up to time when $L\geq \delta L$ for some small $\delta L>0$ chosen to avoid numerical error and instability.
The equation \eqref{eq:intstar} is supplemented with the vertex condition:
\begin{subequations}\label{eq:bdrystar}
	 \begin{equation}
		 \hat{c}(\cdot,t)\text{ is continuous on }\Gamma_*,
	 \end{equation}
	 \begin{equation}\label{eq:vc_star}
		      \sum_{b\in \mathcal{B}^v} \hat{J}_b(v,t)
		      = F_v[\hat{c}],\quad
			v\in \mathcal{\tilde{V}}.
	      \end{equation}
\end{subequations}
Here, $\hat{J}_{b^i_t}:= \hat{a} \hat{c} - \frac{\hat{D}}{L}\partial_{\xi^i} \hat{c}$ and $F_{v^i_*}[\hat{c}(\cdot,t)]=F_{v^i_t}[c]$ takes any of the forms described in this paper.

The setup in \eqref{eq:intstar} and \eqref{eq:bdrystar} is a good candidate for describing a finite difference method.
In the present work, we will only consider the case when all the interior (non-tip) vertices satisfy the Neumann-Kirchoff ($F_v\equiv 0$) vertex condition.
Note that the Neumann-Kirchoff condition applies to both $\hat{J}_{b}$ and $\hat{J}_{b}-\xi \dot{L} \hat{c}$ in \eqref{eq:vc_star} since the $L$ is time invariant on the non-growing edges and we can choose a bond for the edge $e^j_t$ incident to a growing tip so that we have $\xi^j=0$ at $v$.

\AB{add reference to existing FEM}

\subsection{Discretization}
We will describe a numerical method for \eqref{eq:intstar} and \eqref{eq:bdrystar}.
For each $i$, we subdivide $e^i_*$ into $N$ edges by ``adding'' $N-1$ vertices while preserving the direction of each bond $v^{j}_* \overset{b^i_*}{\longrightarrow} v^k_*$ so that it becomes
\begin{equation*}
	v^j_* \sim v^{i,0}_* 
	\overset{b^{i,1}_*}{\longrightarrow} v^{i,1}_* \overset{b^{i,2}_*}{\longrightarrow} v^{i,2}_*\overset{b^{i,3}_*}{\longrightarrow}\cdots \overset{b^{i,N-1}_*}{\longrightarrow}v^{i,N-1}_*\overset{b^{i,N}_*}{\longrightarrow} v^{i,N}_*\sim v^k_*
\end{equation*}
This yields a (combinatorial) graph $\tilde{\Gamma} = (\mathcal{\tilde{V}},\mathcal{\tilde{E}})$ which is a discretization of $\Gamma_*$.
We seek $c^n = (c^n_{v})_{v\in \mathcal{\tilde{V}}}$ such that $c^n_{v^{i,k}_*} \approx \hat{c}(k\delta x, n \delta t)$ on $e^i_*$ where $\delta t>0$ and $\delta x = 1/N$.

Define the backward time difference operator:
\begin{equation}
	(\mathcal{D}^-_t u)^n := \frac{u^n - u^{n-1}}{\delta t}.
\end{equation}
Denote by $\Delta^\top$ the incidence matrix of $\tilde{\Gamma}$.
Let $l^n = (l^n_{e})_{e\in \mathcal{\tilde{E}}}$ and $\dot{l}^n = (\dot{l}^n_{e})_{e\in \mathcal{\tilde{E}}}$ with $l^n_{e^{i,k}_*}     \approx L^i_{n\delta t}$.
For functions on vertices $u^n=(u^n_{v})_{v\in \mathcal{\tilde{V}}}$ and similarly on edges $w^n = (w^n_{e})_{e\in \mathcal{\tilde{E}}}$,
we can define gradient and divergence on $\tilde{\Gamma}$:
\begin{gather}
	\mathrm{grad}_{\tilde{\Gamma}}\, u^n := \left(\mathrm{diag}\frac{1}{l^n\delta x}\right) \Delta u^n,\\
	\mathrm{div}_{\tilde{\Gamma}}\,w^n := -\frac{1}{\delta x}\Delta^\top w^n.
\end{gather}

After choosing averaging operators $\tilde{\mathcal{A}}:\mathbb{R}^\mathcal{V}\to \mathbb{R}^\mathcal{E}$ and $\mathcal{A}:\mathbb{R}^\mathcal{E}\to \mathbb{R}^\mathcal{V}$, we can write the equation for the finite difference scheme:
\begin{equation}
	\begin{split}
		(\mathcal{D}^-_t c)^n + \left(\mathrm{diag} \frac{1}{\mathcal{A}l^n}\right)\mathrm{div}_{\tilde{\Gamma}}\, \left(- \dot{l}^n\tilde{\mathcal{A}}(\tilde{\xi}c^n) + a^n \tilde{\mathcal{A}} c^n - \mathrm{grad}_{\tilde{\Gamma}}\,c^n\right)\\
		+ \mathrm{diag} \left(\mathcal{A} \left(\frac{\dot{l}^n}{l^n}\right)\right) c^n = s^n + g^n
	\end{split}
	\label{eq:disc_c}
\end{equation}
where, in the simple tip growth, we have
\begin{equation}
	g^n_{v^{i,k}_*} =\begin{cases}
		\frac{1}{l^n_{e^{i,k}_*}\delta x} \left(c^n_{v^{i,k}_*} + \kappa^n_{v^{i,k}_*}\right) \dot{l}^n_{v^{i,k}_*} & \text{if $v^{i,k}_*$ is a tip}, \\
		0                                                                                           & \text{otherwise}.
	\end{cases}
\end{equation}
For the setup in \ref{sec:growth_at_junction} where we have coupling between a growing tip and an oscillating network, we simply add similar coupling terms to $g^n$.

To solve this along with the length dynamic, we apply a first order splitting method as in Algorithm~\ref{algo:split}.
\begin{algorithm}
	\caption{First order splitting method}
	\begin{algorithmic}[1]
		\State Initialize $c^0, l^0$
		\For{$n\in \{1,\cdots,n_*\}$}
		\State $\dot{l}^{n}_{e^i_k} = \gamma^{n-1} c^{n-1}_{v^i_N} - \rho^{n-1}$ if $v^i\in \partial\Gamma_*$
		\State $l^n = l^{n-1} + \dot{l}^{n}\delta t$
		\State Solve the linear system \eqref{eq:disc_c} for $c^n$
		\EndFor
	\end{algorithmic}
\label{algo:split}
\end{algorithm}


\end{document}